# Mobile-Aware Scheduling for Low Latency Backhaul over DOCSIS

Jennifer Andreoli-Fang
Office of the CTO and R&D
CableLabs
Boulder, CO

John T. Chapman
Chief Technology and Architecture Office
Cisco Systems
San Jose, CA

*Abstract*—In this paper, we discuss latency reduction techniques for mobile backhaul over Data Over Cable Service Interface Specifications (DOCSIS®) networks. When the latencies from both the wireless and the DOCSIS networks are added together, it can result in noticeable end-to-end system latency, particularly under network congestion. Previously, we proposed a method to improve upstream user-to-mobile core latency by coordinating the LTE and DOCSIS scheduling. The method reduces the impact on system latency from the DOCSIS network's request-grant-data loop, which is the main contributor of backhaul upstream latency. Since the method reduces latency on the DOCSIS data path, it will therefore improve performance of latency sensitive applications, particularly if TCP is used as the transport protocol, especially when the link is congested. In this paper, we investigate the effect of HARQ failure on system performance. Through simulation, we show that despite the uncertainty introduced by the LTE protocol, coordinated scheduling improves overall system latency.

## I. Introduction

The mobile network operators have been under pressure to deploy dense small cell networks in response to the tremendous growth in mobile data usage [1]. All this traffic needs to be backhauled to the mobile core. Fiber has been traditionally preferred by the MNOs to backhaul macrocells. However, fiber is sparse, and to install fiber ubiquitously to backhaul the dense small cell locations is not as economical as using existing fixed infrastructure. Cable operators have built and deployed the hybrid fiber coaxial (HFC) networks everywhere to service broadband residential and commercial customers. This HFC network is an attractive option for small cell backhaul as it is ubiquitous, has ample capacity, and can provide a lower cost alternative to running new fiber.

The HFC network may incur higher latency than the allocated timing budget for mobile backhaul. DOCSIS [2][3] defines the protocol that governs the communication of broadband data between the cable modems (CMs) and the cable modem termination system (CMTS) over the HFC network. The DOCSIS upstream today incurs a minimum latency of 5 ms, average of 11-15 ms, and can be 20 to 50 ms for a loaded network [4].

In comparison, the LTE backhaul budget is lower than the minimum DOCSIS latency [5]. Interference coordination techniques such as CoMP require 5 ms of X2 latency to realize significant gain [6][7]. Furthermore, most of 5G applications require 10 ms user-to-core latency, with some ultra-low latency applications requiring 1 ms latency [8][10]. These requirements are not possible to meet with DOCSIS backhaul today.

Both LTE and HFC networks follow a request-grant-data (REQ-GNT-data) loop for typical upstream transmissions. When wireless data is sent from the user equipment (UE) to the LTE base station (eNB) and backhauled over DOCSIS, the data experiences this type of 3-way loop twice. This occurs when the two networks, the wireless and its backhaul, are uncoordinated.

In previous papers [16][17], we proposed a method to improve the upstream UE-to-mobile core latency by coordinating the LTE and DOCSIS scheduling operations. A 2-stage pipeline is formed with the LTE and DOCSIS schedulers. In this paper, we investigate details such as the effect of HARQ failure on latency reduction. We found that the system latency is significantly reduced despite the uncertainties introduced by variables such as HARQ failure.

This paper is organized as follows. Section 2 provides a brief introduction on the DOCSIS and LTE upstream scheduling operations. Section 3 discusses the proposed method in detail. Section 4 describes an end-to-end MATLAB simulator environment including DOCSIS and LTE components developed to investigate the benefit of coordinated scheduling operations. Simulation results that show latency improvement are also described in the section.

## II. Background and Previous Work

### A. DOCSIS Upstream Data Plane Latency

The DOCSIS protocol specifies several major types of upstream scheduling services including best effort (BE), real-time polling service (RTPS), and unsolicited grant service (UGS). Most upstream data is transmitted with BE service. Best effort scheduling follows a request-grant-data loop as shown in Fig. 1 (also see [4]). The requests are sent in the contention regions which the CMTS schedules regularly.

RTPS was designed to support real-time data flows that generate variable size packets periodically where the CMTS provides unicast request opportunities periodically. As shown in Fig. 1, after the CM detects data arrival and formulates a bandwidth request (REQ), it waits either for a contention region or a polling opportunity to transmit the REQ, depending on the scheduling service the traffic is configured for.

UGS was designed to support real-time data flows, such as VoIP, that periodically generate fixed size packets. The CMTS provides fixed-size grants of bandwidth on a periodic basis. The CM utilizes the periodic grants to transmit data directly without sending REQs.

The CMTS scheduler typically processes REQs every 2 ms and generates a MAP that describes 2 ms worth of grant allocations. Since the CMTS sends at least one MAP in advance of that MAP's allocation start time, the shortest REQ-GNT cycle on DOCSIS is theoretically 4 ms. Additionally, the CM and the CMTS each need processing and lead time, typically 0.5 ms on each device. So the practical minimum upstream latency for DOCSIS is about 5 ms.

The average latency is expected to be higher due to the random arrivals of data and therefore, REQs, as well as the scheduling of contention regions. Higher network loading may also result in collision of the REQs, which triggers a truncated exponential backoff that will further increase latency. The average DOCSIS upstream latency for best effort traffic has been measured to be 11-15 ms with the potential of a significantly higher maximum latency. For further discussion on this topic, see [4].

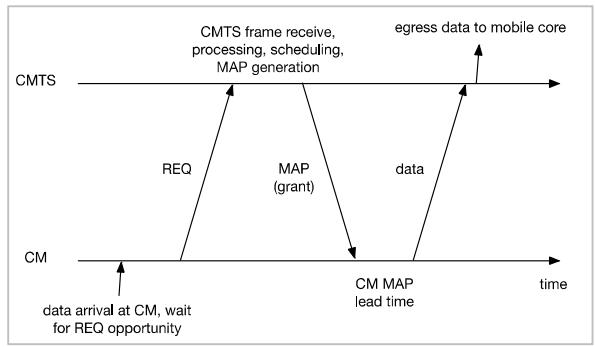

Fig. 1. DOCSIS REG-GNT-data loop for BE or RTPS scheduling service

### B. LTE Uplink Data Plane Latency

LTE uplink (UL) access follows a similar 3-way REQ-GNT-data loop but with some differences. Referring to Fig. 2a, which shows the signaling between UE and eNB, when data arrives at the UE, the UE first determines if it has a valid LTE UL grant. If it does not, the UE waits for a Scheduling Request (SR) opportunity.

Upon receiving the SR, the eNB schedules an UL grant for the UE to transmit a Buffer Status Report (BSR). The eNB turnaround time is 4 ms. Once the BSR is received by the eNB, it schedules an UL grant for the UE to transmit data. The typical minimum LTE uplink latency for transmitting one short packet with no network loading is summarized in Table I.

Note that the 21 ms average minimum latency for LTE is greater than the 5 ms minimum or 11-15 ms average latency for DOCSIS.

### C. Sum it Altogether

Referring again to Fig. 2a, when data arrives at the UE to be transferred uplink to the eNB, and backhauled on the DOCSIS link to the mobile core, it needs to traverse two 3-way REQ-GNT-data loops on LTE and on DOCSIS. First, the UE makes a scheduling request which is followed by an LTE grant that allows the data to traverse through the LTE air interface. Once the data reaches the eNB and is egressed to the CM, a DOCSIS request is made by the CM to the CMTS. The CMTS then assigns DOCSIS grant(s) which allow the data to traverse through the DOCSIS backhaul where it eventually reaches the mobile core. The LTE and DOCSIS latencies are added in serial.

TABLE I. LTE UPLINK LATENCY COMPONENTS

| Latency components | Latency (ms) |
|---|---|
| Waiting time for SR (assume configured SR period of 5 ms) | 0.5 – 5.5 |
| UE sends SR, eNB decodes SR, eNB generates grant for BSR | 4 |
| eNB sends grant, UE processes grant, UE generates BSR | 4 |
| eNB processes BSR, eNB generates grant for data | 4 |
| eNB sends grant, UE processes grant, UE sends UL data | 4 |
| eNB decodes UL data (estimate) | 1.5 – 2.5 |
| Total | 18 – 24 |

### D. Previous Work

To the best of our knowledge, there has been no previous attempt to reduce the latency by coordinating wireless and backhaul networks. Efforts have been made to separately reduce the LTE latency and the DOCSIS latency. The 3GPP is working on techniques including contention based data transmissions, and reducing the number of transmission time intervals [11]. However, the proposed techniques have yet to be standardized into the 3GPP Stage 2 specifications.

There has not been a driving need to reduce the DOCSIS latency, except for Active Queue Management (AQM) [12]. The AQM feature applies when multiple applications share a network connection and requires fundamental modem hardware change. One CMTS vendor is innovating with implementing faster MAC scheduler and predictive scheduler, which proactively grants in an intelligent manner [4].

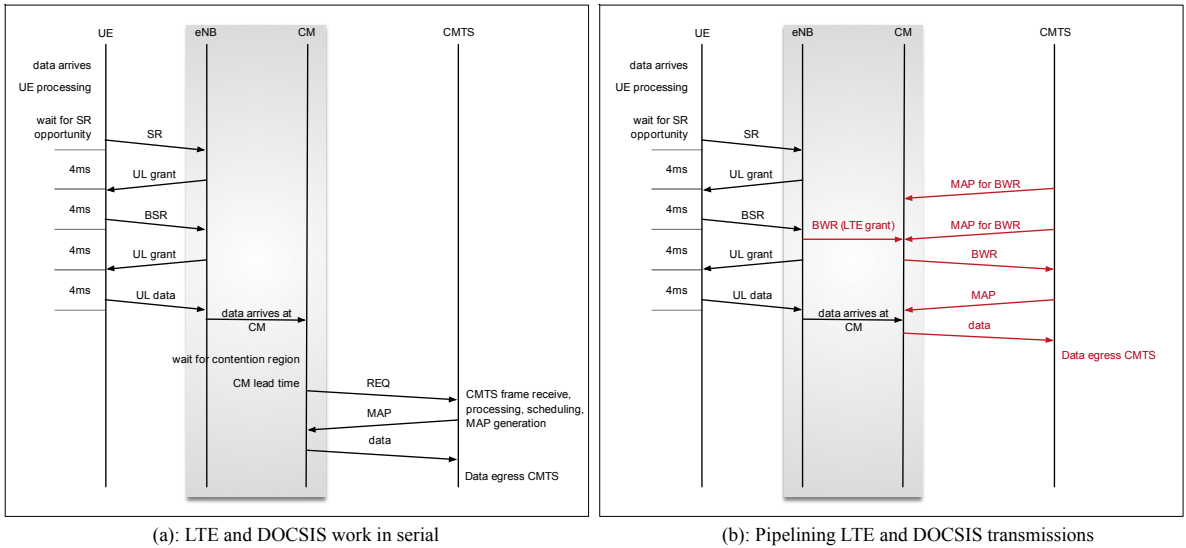

(a): LTE and DOCSIS work in serial

(b): Pipelining LTE and DOCSIS transmissions

Fig. 2. Backhauling LTE Data with a DOCSIS Network

The work described in this paper differ from all these efforts. Our method can be implemented in an API into the DOCSIS scheduler. This API allows the information on expected future data transmission from LTE to be sent to the DOCSIS scheduler. In this way, the method proposed here works with any existing CMTS schedulers and requires minimal change to the software.

III. PIPELINING THE LTE AND DOCSIS SCHEDULING OPERATIONS

This section proposes a method to improve the upstream latency by coordinating the LTE and DOCSIS scheduling loops. Since the DOCSIS REQ-GNT-data loop is the main contributor of backhaul upstream latency, the REQ-GNT-data processes on LTE and DOCSIS links are pipelined to reduce latency.

The first stage of the pipeline consists of the LTE REQ-GNT-data operation. Instead of waiting for the LTE data to arrive at the eNB, the eNB indicates to the CM that it is expecting UL data in the near future. This information prompts the second stage of the pipeline, the DOCSIS 3-way loop to start early. The DOCSIS scheduler prepares a grant in advance, just in time to transport the LTE data when it eventually arrives at the CM.

Effectively, the DOCSIS activities of sending a DOCSIS REQ on the CM, and receiving, scheduling, MAP generation activities on the CMTS occur in parallel to the activities taking place on the wireless link. In this way, the operation through the two networks become pipelined, where the first pipeline stage is the LTE scheduler, which informs the next stage, the DOCSIS scheduler, of what is about to come.

The details of the pipelining operations is shown in Fig. 2b. After the eNB has received the BSR and generated an LTE UL grant, the eNB MAC generates a Bandwidth Report (BWR) message, indicating the amount of grant and the grant time. This message is forwarded to the CM. Generally, the eNB and the CM are connected via a Gigabit Ethernet connection and the propagation time of the message is negligible.

Meanwhile, on the DOCSIS side, the CMTS periodically allows the CM to transmit the BWR message by sending an upstream (US) grant allocation MAP. Since the eNB scheduler generates UL data allocation every subframe, the granting interval on the DOCSIS link can also be every 1 ms. The CMTS scheduler uses the information in the BWR message to generate an US grant including the scheduled transmission start time to the CM to transport UE data. When LTE UL data egresses the eNB and arrives at the CM, a DOCSIS US grant has arrived and is waiting to be used at the CM to immediately forward the LTE data to the CMTS.

A. The Bandwidth Report

The BWR needs to be sent to the CMTS quickly in order for the CMTS scheduler to make use of the time sensitive information and to generate an US grant. The CMTS can utilize either UGS or RTPS scheduling service to allow the CM to skip the contention request process to accelerate the transport of the BWR messages.

With RTPS, periodic polling grants for REQs are issued to the CM. When a BWR message arrives at the CM, the CM indicates the size of the message when it is polled. The CMTS then issues a US grant to transport the actual message. With UGS, the CM uses periodic pre-allocated UGS grants to transport the BWR messages. Either method needs to transport BWR messages in time – the messages need to arrive in time to be used by the CMTS scheduler approximately 4 ms before the expected egress time of the UE data. While UGS allows the BWR to arrive at the CMTS sooner compared to RTPS, UGS may waste more DOCSIS grants.

The BWR may describe the LTE grant as a "bulk grant," indicating a single block of bytes that the eNB has allocated for all UE transmissions in a future subframe.

Alternatively, to allow for traffic differentiation on the DOCSIS link, the BWR can report the amount of UL grant for each Logical Channel Group (LCG). Each UE's BSR includes the number of bytes requested for each of the UE's 4 LCGs. This information helps the eNB MAC scheduler determine the number of bytes to be granted for each LCG, even though the UL grant is only expressed in terms of a single number of bytes. The eNB MAC includes 4 blocks of bytes in the BWR, with each block indicating the aggregate number of bytes allocated for an LCG for all the UEs scheduled to transmit in the future subframe described by the BWR. Because each LCG maps to a different set of QoS parameters, the CMTS scheduler uses this information to prioritize the LTE data traffic in a more granular manner.

The BWR takes up a fraction of the DOCSIS upstream capacity. For an 80-byte length BWR, if a BWR is sent every 1 ms, then the BWR will consume 640 kbps per attached small cell.

### B. Effect of HARQ Failures on BWR Computation

LTE uses Hybrid Automatic Repeat reQuest (HARQ) to increase reliability and reduce latency associated with the air interface transmission failures. Retransmissions occur at fixed 8 subframe intervals after the initial transmission has failed, in frequency division duplex (FDD) mode.

To enable higher throughput, an LTE eNB maintains 8 UL HARQ processes for each UE. When data sent using a HARQ process fails the CRC test, the process becomes inactive and cannot be reused for new data, until the original failed data is scheduled to retransmit 8 subframes later. Meanwhile, the UE will continue to send UL data using other active HARQ processes. This new data, if passes the CRC tests, will be queued at the eNB's Radio Link Control (RLC) layer. The new data will egress the eNB when the earlier failed data is retransmitted 8 subframes later and succeeds the HARQ process.

Assuming the current time is when the eNB just detected a HARQ failure. The BWR to be egressed at current time needs to include the expected new data that will arrive in the 8 future subframes. Based on this, the CMTS generates just-in-time grants that will allow the original and the new data to be transported on the DOCSIS link as soon as it together egresses the eNB and arrives at the CM. The amount of future data expected to arrive in the next 8 subframes is known to the eNB, because each arrival has been scheduled in a sliding time window of 8 subframes ago.

Denote $ULGrant(t)$ as the UL grant that the eNB scheduler has computed for a UE, and $BWR(t)$ as the BWR for time $t$. The BWR is calculated as

$$BWR(t) = \begin{cases} ULGrant(t), & \text{HARQ succeeds at } t \\ ULGrant(t) + \sum_{x=t-8}^{t} ULGrant(x), & \text{HARQ fails at } t \end{cases}$$

Let us look at a simple example shown in Fig. 3. We assume that the first LTE UL data request is received some time before subframe -8, such that the first UL scheduling grant is generated at subframe -8. The eNB MAC scheduler grants $ULGrant(-8) = TB0$ at subframe -8, and expects the granted data to be received at subframe 0. If TB0 fails the CRC test, it will be retransmitted at subframe 8. The BWR that the eNB issues at subframe 0 is then $BWR(0) = ULGrant(0) + TB0 + TB1 + \cdots + TB7$. Note that the eNB must keep track of the amount of grants for at least the previous 8 subframes.

## IV. SIMULATION AND RESULTS

In this section, we present results from a custom MATLAB simulator to analyze the latency performance of the baseline system and the proposed method. This unique MATLAB model contains models for the UE, the eNB, the CM and the CMTS.

Simulation assumptions are summarized in Table II. The simulation platform models the end-to-end LTE and DOCSIS system with protocol signaling capability relevant for this study. In the LTE stack, the Media Access Control (MAC) and Radio Link Control (RLC) layers are modeled in full with some PHY layer abstractions. In the simulation setup, multiple UEs are connected to a single eNB, and multiple UEs are capable of transmitting upstream traffic. Backhaul is provided by a CM communicating with a CMTS.

In this study, the BWR is constructed by the eNB every 2 ms. The message is then forwarded to the CM and is sent using the UGS scheduling service on the DOCSIS network. The frequency of the BWR transmission by the eNB is designed to match the periodicity of UGS grant as set by the CMTS which is 2 ms in this study. This can be reduced to 1 ms as needed. LTE data is backhauled on the DOCSIS network by using a best effort (BE) scheduling service for the baseline, and by using scheduled grants when the BWR is used.

The LTE deployment is FDD with a system bandwidth of 10 MHz for both uplink and downlink. It is assumed that the LTE system uses SISO (Single-Input

| | | | | | | | HARQ success | | | HARQ retx success | | | HARQ failure | | | No bytes transmitted on air | | | |
|---|---|---|---|---|---|---|---|---|---|---|---|---|---|---|---|---|---|---|---|
| BWR(t) | BWR(-8) | BWR(-7) | BWR(-6) | BWR(-5) | BWR(-4) | BWR(-3) | BWR(-2) | BWR(-1) | BWR(0) | BWR(1) | BWR(2) | BWR(3) | BWR(4) | BWR(5) | BWR(6) | BWR(7) | BWR(8) |
| t | -8 | -7 | -6 | -5 | -4 | -3 | -2 | -1 | 0 | 1 | 2 | 3 | 4 | 5 | 6 | 7 | 8 |
| eNB expects granted bytes at t = | 0 | 1 | 2 | 3 | 4 | 5 | 6 | 7 | 8 | 9 | 10 | 11 | 12 | 13 | 14 | 15 | 16 |
| Granted amount | TB0 | TB1 | TB2 | TB3 | TB4 | TB5 | TB6 | TB7 | ULGrant(0) | ULGrant(1) | ULGrant(2) | ULGrant(3) | ULGrant(4) | ULGrant(5) | ULGrant(6) | ULGrant(7) | ULGrant(8) |
| Bytes received | 0 | 0 | 0 | 0 | 0 | 0 | 0 | 0 | 0 | TB1 | TB2 | TB3 | TB4 | TB5 | TB6 | TB7 | TB0 |
| Bytes egressed | 0 | 0 | 0 | 0 | 0 | 0 | 0 | 0 | 0 | 0 | 0 | 0 | 0 | 0 | 0 | 0 | TB0 +…+ TB7 |
| Bytes transmitted on air | 0 | 0 | 0 | 0 | 0 | 0 | 0 | 0 | TB0 | TB1 | TB2 | TB3 | TB4 | TB5 | TB6 | TB7 | TB0 |

Fig. 3. Example BWR Computation Due to HARQ Blocking

Single-Output) transmission mode. The BSR is transmitted by the UE with a periodicity of 10 ms. HARQ processes are modeled for both bursty traffic such as live video streaming and periodic traffic such as in VoIP traffic. The LTE eNB employs a round-robin scheduler when multiple UEs are present in the system. A slow fading wireless channel model is assumed in which the channel stays static for the duration of 10 ms. The physical channel is modeled by varying a UE's MCS (modulation and coding scheme) such that each UE's MCS has an average MCS index of 22 with a normal distribution between MCS indices 18 and 26.

TABLE II. SIMULATION SETTINGS

| Simulation parameters | |
|---|---|
| Simulation time | 2 seconds (2000 subframes) |
| **DOCSIS parameters** | |
| Number of CMTS and CMs | 1 CMTS, 1 CM |
| Scheduling service for BWR | UGS |
| BWR periodicity | 2 ms |
| UGS grant periodicity | 2 ms |
| Scheduling service for data | Best effort |
| **LTE system parameters** | |
| Number of eNBs | 1 or 4 |
| Number of UEs per eNB | 1 or 6 |
| Duplexing method | FDD |
| System bandwidth | 10 MHz |
| Spatial multiplexing | SISO |
| BSR periodicity | 10 ms |
| Channel | Slow fading with channel updates periodic of 10 ms (MCS 18-26) |
| HARQ | OFF or ON (10 % BLER) |
| **LTE eNB parameters** | |
| Scheduler type | Round robin |
| BWR periodicity | 2 ms |
| **Traffic parameters** | |
| Upstream traffic type | Case 1: Facebook Live video streaming / upload Case 2: uniform traffic, 200 byte packets @ 1 ms inter-arrival |

In the simulations, we study latency reduction comparing 2 BWR implementations.

We first consider the performance of the baseline scenarios where no coordination exists between the DOCSIS and the LTE networks. We then study the proposed BWR method where the eNB signals the expected future LTE transmission information to the DOCSIS system which enables just-in-time scheduling grants to be made by the CMTS. The latency of the end-to-end uplink transmission is measured in the simulations from the time a packet arrives at the UE to the time it egresses the network side of the CMTS. The measurement is sampled at the input and output of MAC layers in each side.

The goal of this scenario is to show the effectiveness of computing the BWR taking into account the head of the egress queue blocking due to HARQ failure, as discussed in Section 3B. There are two different ways to implement the BWR:

1. BWR without buffer flush: this is the UL grant-only BWR, which includes only the UL grant computed by the eNB MAC scheduler for the subframe. Implementing BWR this way requires the interaction with the eNB MAC scheduler.

2. BWR with buffer flush, which, in addition to the UL grant-only BWR, includes the expected new data that will arrive in the 8 future subframes. This implementation requires the interaction with eNB HARQ processing, in addition to the scheduler.

When a HARQ failure occurs, the new data is held in the eNB RLC layer, and will miss the just-in-time DOCSIS grants. With implementation #1, BWR without buffer flush, the waiting DOCSIS grant can only accommodate the original retransmitted data. The CM will need to revert to the normal DOCSIS REQ-GNT mechanism to request grants for the new data. The new data will eventually be transported on the DOCSIS link, but without the benefit of latency reduction provided by the BWR. With implementation #2, BWR with buffer flush, when the new and the original data egress eNB and arrive at the CM after a successful retransmission, there will be a DOCSIS just-in-time grant large enough to accommodate it altogether.

To implement BWR with buffer flush, the eNB must also keep track of the amount of grants for at least the previous 8 subframes. State keeping, as well as interacting with other protocol layers increases system complexity. In this scenario, we want to investigate whether the additional complexity is worth the gain in latency reduction.

We keep the scenario simple to focus on comparing the gain in latency reduction between the 2 BWR implementations. In this scenario, the CMTS serves a CM, which backhauls traffic for 1 eNB that serves 1 UE. The UE sends 200-byte packets at 1 ms fixed interval.

Fig. 4 compares the latency CDF of the end-to-end system, for baseline (no BWR), BWR without buffer flush, and BWR with buffer flush, averaged over all packets transmitted by the UE. The latency gain of the BWR with buffer flush compared to without is 0.65 ms averaged over all packets. Since the HARQ failures affect only a portion of the traffic, Table III shows the DOCSIS and the system latency for the affected packets only, implementing BWR with and without buffer flush. We observe the latency reduction gain of with buffer flush compared to without is 2 ms minimum, 2.6 ms average, and 4 ms maximum for the DOCSIS link.

Note that since the system setup does not include contention, the maximum gain is limited by the 4 ms maximum. Because the background traffic on the DOCSIS link is not modeled, the DOCSIS link incurs the theoretical minimum of 5 ms of latency. The purpose of BWR is to "hide" all or part of DOCSIS REQ-GNT cycle under the LTE latency. We observe that the BWR has eliminated all the possible DOCSIS REQ-GNT loop, which is 4 ms. What is left is the latency of the managed queue in the CM, which includes framing and error correction.

## V. CONCLUSION

In this paper, we present a method to improve the upstream system latency for a mobile network backhauled over a DOCSIS network. The operations of the two networks are pipelined. The first stage of the pipeline is the eNB scheduler: it constructs Bandwidth Reports based on its scheduling information. The BWR informs the second stage of the pipeline, the DOCSIS scheduler, of the traffic that will need to be backhauled in the near future. This enables the DOCSIS scheduler to proactively send just-in-time grants to the modem. Compared to non-coordinated system, the BWR-based method show sizeable latency reduction that becomes significant as the system becomes loaded. We also present 2 implementations of the BWR, and show that additional implementation complexity is worth the gain in latency reduction.

## ACKNOWLEDGMENT

The LTE and DOCSIS simulator environment was built by the following engineers: Michel Chauvin of CableLabs, and Elias Chavarria Reyes and Dantong Liu of Cisco. Many engineers provided technical insights throughout this project. The lead authors from CableLabs and Cisco would like to thank Joey Padden, Balkan Kecicioglu and Vaibhav Singh of CableLabs, and Oliver Bull, Alon Bernstein and Zheng Lu of Cisco for the technical discussions.

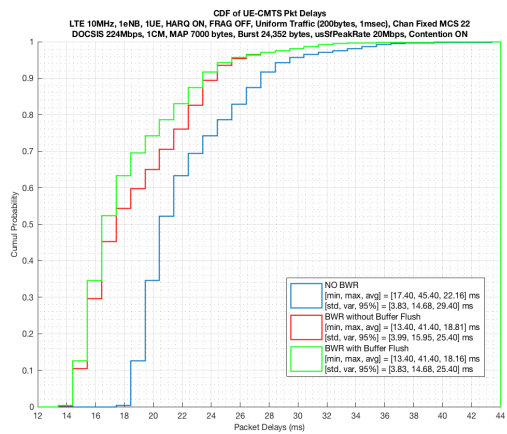

Fig 6. Latency CDF of the System, Baseline vs. BWR Without Buffer Flush vs. BWR With Buffer Flush

TABLE III. LATENCY (IN MILLISECONDS) FOR PACKETS AFFECTED BY HARQ FAILURE (SCENARIO 2)

| | End to end (LTE+DOCSIS) | | | DOCSIS only | | |
|---|---|---|---|---|---|---|
| | Min | Avg | Max | Min | Avg | Max |
| BWR without buffer flush | 15.4 | 20.7 | 27.4 | 3.2 | 4.1 | 6.2 |
| BWR with buffer flush | 13.4 | 18.2 | 25.4 | 1.2 | 1.5 | 2.2 |